\documentclass[11pt,aps,pra,reprint,superscriptaddress]{revtex4-2}
\usepackage{graphicx}
\usepackage{hyperref}
\usepackage{braket}
\usepackage{amsmath}
\usepackage{amssymb}
\usepackage{mathtools}
\usepackage{dsfont}
\usepackage{systeme}
\usepackage{titlesec}
\usepackage{bbm}
\usepackage{xcolor}

\begin{document}

\title{Locating quantum critical points with shallow quantum circuits}

\author{Zhi-Quan Shi}
\affiliation{Guangdong Provincial Key Laboratory of Quantum Engineering and Quantum Materials, School of Physics and Telecommunication Engineering, South China Normal University, Guangzhou 510006, China}

\author{Fang-Gang Duan}
\affiliation{Guangdong Provincial Key Laboratory of Quantum Engineering and Quantum Materials, School of Physics and Telecommunication Engineering, South China Normal University, Guangzhou 510006, China}

\author{Dan-Bo Zhang}
\email{dbzhang@m.scnu.edu.cn}
\affiliation{Guangdong-Hong Kong Joint Laboratory of Quantum Matter, Frontier Research Institute for Physics, South China Normal University, Guangzhou 510006,
	China}
\affiliation{Guangdong Provincial Key Laboratory of Quantum Engineering and Quantum Materials, School of Physics and Telecommunication Engineering, South China Normal University, Guangzhou 510006, China}

\date{\today}

\begin{abstract}
Quantum critical point is one key concept for studying many-body physics but its investigation may be resource-demanding even for a quantum computer due to the intrinsic complexity. In this paper, we propose an approach based on variational quantum eigensolver(VQE), dubbed as Delta-VQE, for locating the quantum critical point using only shallow quantum circuits. With Delta-VQE, the critical point can be identified as a most confusing point, quantified as zero difference between two variational energies that use two representative reference states of distinct phases of matter. Remarkably, the signature of a critical point as a minimal point can be sharper while using shallower quantum circuits. We test the algorithm for different quantum systems and demonstrate the usefulness of Delta-VQE. The scheme suggests a new avenue for investigating quantum phases of matter on near-term quantum devices with limited quantum resources.     
\end{abstract}

\maketitle
\section{Introduction} \label{sec:introduction}
Understanding quantum phases of matter is one key topic in studying quantum many-body physics. 
Quantum critical point~\cite{Sachdev}, which separates different phases of matter, becomes of special interest 
for its novel physical properties, e.g., divergence of the correlation length~\cite{Sachdev} and violation of the area law of the entanglement entropy~\cite{Vidal_PRL_2003,Schuch_PhysRevLett_2008,Eisert_RevModPhys_2010}. For fully describing a quantum critical point, it is desirable to solve the ground state of the underlying quantum system, which is known to be hard for a classical computer due to generic fermionic sign-problem~\cite{Feynman_1982,troyer_05}. On the other hand, simulating a quantum critical point with a quantum computer can be efficient~\cite{Poulin2009PreparingGS,Ho_SciPostPhys_2019,Zhu2020GenerationOT,Zhang2021ContinuousVariableAT}. However, it in general demands more quantum resources due to the intrinsic complexity of describing the critical point with a quantum circuit, even with variational quantum algorithms that are designed for implementing in near-term quantum devices~\cite{Ho_SciPostPhys_2019,Zhu2020GenerationOT}.  

However, the task of locating a quantum critical point rather than revealing its properties can be easier and solving the ground state faithfully may be not necessary. The situation is closely related to classification in machine learning, in which partial information may lead to correct answers~\cite{trevor_hastie_elements_2009,Goodfellow-et-al-2016}. There are some approaches of machine learning for studying phases of matter~\cite{wanglei_PRB_2016,Juan_2017NatPhys,Nieuwenburg2017LearningPT}. With only some descriptors of the quantum system, e.g., correlation functions or entanglement spectrum, different phases of matter can be recognized and therein quantum critical points stand out. Remarkably, with a confusion scheme in unsupervised machine learning, a quantum critical point can be identified as an extreme point of performance for machine learning~\cite{Nieuwenburg2017LearningPT}. However, those machine learning approaches still require data that is ultimately based on available ground state, such as measurement on real materials or quantum simulators, or from numeral simulations, e.g., using neural networks~\cite{Carleo2017SolvingTQ}. A question can be asked: can we locate a quantum critical point without truly solving the ground state of a quantum system? 

To answer this question, it is inspiring to consider how we can identify if one object belongs to two classes of distinct objects in real life or in physics. Obviously, it is not necessary to dig into all details. Rather, one may roughly compare the object with two representative distinct objects and get the answer from the similarity. For identifying quantum phases of matter, the quantum critical point is special as it is unlike either phase. In this regard, the quantum critical point can be most confusing for identification as to which phase. This inspires us to locate the quantum critical point by comparison, but without solving the ground state accurately. 

In this work, we propose a variational quantum algorithm, called Delta-VQE, that can locate quantum critical points with only shallow circuits. The Delta-VQE compares two variational energies with reference states corresponding to distinct phases of matter. Although the quantum eigensolver can not solve the ground state energy accurately with a limited depth of quantum circuit, the difference of two energies on the quantum critical point can be minimal and thus can be regarded as a signature for locating the critical point. The Delta-VQE algorithm is tested on several typical models involving different types of quantum phase transitions. Numeral simulations suggest that shallower circuits can lead to sharper minimalism.  
Our work suggests an avenue for studying quantum phases of matter on near-term quantum computers. 

The paper is organized as follows. We first give motivations and introduce the Delta-VQE algorithm in Sec.~\ref{sec:introduction}. Then, we test the algorithm on several typical models with numeral simulations in Sec.~\ref{sec:results}. Finally, we give   conclusions in Sec.~\ref{sec:conclusion}.

\section{The quantum algorithm framework}\label{sec:delta-vqe}
In this section, we first introduce some backgrounds on quantum critical points and variational quantum algorithms and then give the motivation behind the variational quantum algorithm for locating the quantum critical point. The Delta-VQE is proposed with detailed procedures. 

\subsection{Quantum critical point and variational quantum eigensolver}
For a quantum system described by a Hamiltonian $H(h) $with an intrinsic parameter  $h$, a quantum phase transition occurs when the gap~(the energy difference between the first excited state and the ground state) closes at a parameter $h_c$ when varying $h$~\cite{Sachdev}. The quantum phase transition point is also called quantum critical point, for the reason that physical properties close to the phase transition show critical behaviors, e.g., the correlation length becomes divergent. The quantum critical point can separate two different phases of matter in the phase diagram. For this reason, identifying quantum critical points can be useful for obtaining the phase diagram. 

One may solve a group of Hamiltonian $\{H(h)\}$ with varying $h$ and identify the critical point as a gapless point. For each Hamiltonian, solving its ground state is a typical quantum many-body problem and there is an intrinsic difficulty for generic quantum systems with classical algorithms~\cite{troyer_05}. Remarkably, solving the quantum critical point or the gapless quantum system can be more difficult than gapped ones. One great promise made by quantum computing is that simulations of quantum systems can be efficient~\cite{Feynman_1982,nielsen_chuang_2010}. Solving ground state is of special interest and there are different approaches of quantum algorithms, such as quantum phase estimation~\cite{Poulin2009PreparingGS}, adiabatic quantum algorithm~\cite{Farhi2000QuantumCB,Albash_RevModPhys_2018}, variational quantum eigensolver~\cite{MYung2014, McClean2016,VQA-Cerezo_2021}. Among them, VQE is selective for implementation on near-term quantum computers, as it can use affordable quantum resources by targeting the desired ground state with a parameterized quantum circuit of moderate depth. Remarkably, it can be argued that preparing the ground state for a quantum critical point requires only a polynomial depth of the system size~\cite{Ho_SciPostPhys_2019}.

The behavior and performance of variational quantum eigensolver strongly depend on the ansatz. For our purpose, we adopt a special ansatz, called Hamiltonian variational ansatz~(HVA)~\cite{wiersema-PRXQuantum2020})~(also closely related to the quantum alternating operator ansatz~\cite{farhi2014quantum,Hadfield_QAOA_2019}), that can inherit symmetries of Hamiltonian in the ansatz~\cite{Li:2021kcs} and also is closely related to the adiabatic quantum algorithm~(quantum annealing)~\cite{Farhi2000QuantumCB,farhi2014quantum}. 

Let us write the Hamiltonian as a summation of local Hamiltonians encoded with qubits, $H(h)=\sum_{i=1}^{N} c_i(h) L_i$, where $L_i$ is a product of Pauli matrices.
The VQE aims to solve the ground state for a Hamiltonian with a given $h$. The  ansatz of the ground state can be expressed as,
 \[
 \ket{\psi(\boldsymbol{\theta})}=U(\boldsymbol{\theta})\ket{\psi_0},
 \]
where $\ket{\psi_0}$ is a reference state and $U(\boldsymbol{\theta})$ is an unitary operator with a set of parameters $\boldsymbol{\theta}$. The optimized parameters $\boldsymbol{\theta}$ should be obtained by minimizing the variational energy
\begin{equation}\label{eq:ene}
	E(\boldsymbol{\theta};h)=\bra{\psi(\boldsymbol{\theta})}H(h)\ket{\psi(\boldsymbol{\theta})}.
\end{equation}
This can be achieved with a hybrid quantum-classical computing by evaluating the energy on a quantum processor and updating the parameters with classical computing~\cite{McClean2016}.     

We adopt the Hamiltonian variational ansatz which constructs the unitary evolution $U(\boldsymbol{\theta})$ using alternative Hamiltonian evolutions with non-commuting Hamiltonians~\cite{wiersema-PRXQuantum2020,Li:2021kcs}. For this, the Hamiltonian is divided as a summation of terms $H=c_1H_1+c_2H_2+\cdots + c_nH_n$ with $n\ge2$, where $[H_i,H_j]\not=0$ but all terms in one $H_i$ commute. The $U(\boldsymbol{\theta})$ of multiple layers can be constructed by evolving $H_j$ with time duration $\theta_{ij}$ in the $i$-th layer,
\begin{equation} \label{eq:HVA}
U(\boldsymbol{\theta}) \equiv \prod_{i=1}^p \prod_{j=1}^n
	\exp(i\,\theta_{ij}H_j),
\end{equation}
where $p$ is the number of layers and larger $p$ will render the parameterized quantum circuit more expressive. By construction, the unitary operator $U(\boldsymbol{\theta})$ will have the same symmetry as the Hamiltonian. 

The Hamiltonian variational ansatz is closely related to the quantum adiabatic algorithm by viewing it as a trotterization of the adiabatic process~\cite{farhi2014quantum,Chandarana_PhysRevResearch_2022}.  At the $p\rightarrow\infty$ limit, the Hamiltonian variational ansatz can approach the quantum adiabatic process with proper variational parameters~\cite{farhi2014quantum}. The Hamiltonian variational ansatz can have an advantage over the quantum adiabatic algorithm since it may use a shallower circuit by optimizing each evolution time period in Eq.~\eqref{eq:HVA}.

For the Hamiltonian variational ansatz, the reference state is typically chosen as the ground state for a term $H_i$ in the Hamiltonian $H$, which can be a product state. Compared with the ground state $\ket{\psi_0}$ of $H_i$,  the ground state of $H$ should add fluctuations on $\ket{\psi_0}$, which is the goal of $U(\boldsymbol{\theta})$ with proper parameters $\boldsymbol{\theta}$. 

\subsection{Delta-VQE}
The Hamiltonian variational ansatz is suggested to be efficient for solving ground states of quantum many-body systems~\cite{wiersema-PRXQuantum2020}.  For instance, it requires $p=O(N/2)$ to solve the critical point of the transverse field quantum Ising model~\cite{Ho_SciPostPhys_2019,wiersema-PRXQuantum2020}. For generic quantum systems, the time complexity will be larger but the required layers of HVA may be polynomial of the system size for a desirable accuracy $\epsilon$~\cite{wiersema-PRXQuantum2020}. While locating the quantum critical point may be achieved by solving the system accurately on a quantum computer, the requirement of quantum resources is still huge for the current NISQ quantum devices. It wonders if we can exploit some intrinsic features of quantum phases of matter that locating the quantum critical point can be realized with shallow quantum circuits. 

It is inspiring to first review the quantum adiabatic algorithm~(QAA)~\cite{Albash_RevModPhys_2018}. The basic idea of QAA is to start with a Hamiltonian whose ground state is easy to prepare. Then tuning the Hamiltonian into the target Hamiltonian adiabatically, the ground state of the target Hamiltonian can be naturally reached. To fulfill the adiabatic condition, the gap of each Hamiltonian along the adiabatic path should not be closed.  Let us denote $\ket{\psi_0}$ and $\ket{\psi_1}$ describe two different phases at $h<h_c$ and $h>h_c$, separately. Then, to prepare a ground state  for $H(h)$ with $h<h_c$~($h>h_c$) adiabatically, it should start with $\ket{\psi_0}$~($\ket{\psi_1}$). Thus, to make the QAA work for investigating quantum phases of matter, it is important to make sure that the initial state and the target state belong to the same phase so that crossing the gapless point can be avoided. Remarkably, the quantum critical point is special since either starting with $\ket{\psi_0}$ and $\ket{\psi_1}$ can not avoid the issue of closing of the gap, and thus preparing a critical state can not be accurate.
In this regard, there is an asymmetry for the QAA for preparing a ground state using two initial states belonging to different phases of matter in general, while this asymmetry disappears for the critical point. 

\begin{figure}
	\centering
	\includegraphics[width=1\linewidth]{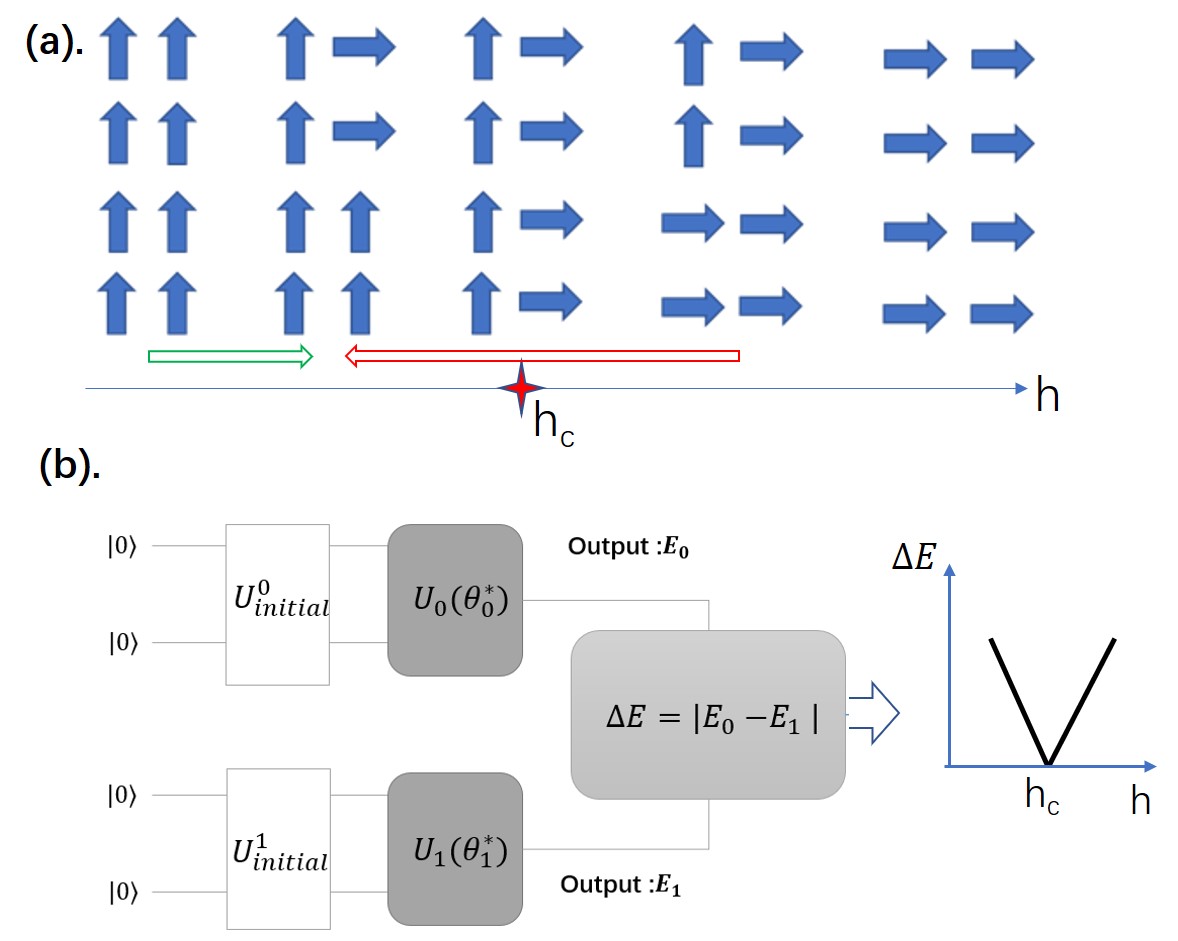}
	\caption{Illustration of the Delta-VQE algorithm. (a). Illustration of different phases with varying $h$ and $h_c$ represents the quantum critical point. To evolve into a given state from an initial state, it is harder when the evolution should get across the critical point~(red arrow) than one~(green arrow) that does not. (2). A pictorial representation of the Delta-VQE algorithm, which compares two optimized variational energies that using two initial states corresponding to distinct phases of matter. Here $U^0_\text{initial}$ and $U^0_\text{initial}$ are used to prepare two initial or reference states. The critical point can be located as the minimum of the absolute energy difference.}
	\label{fig:deltaalgorithm}
\end{figure}

The asymmetry of QAA can inspire the design of variational quantum algorithms for detecting the quantum critical point. Due to the close relation between QAA and HVA, we adopt the ansatz of HVA. Unlike QAA which requires a long time for evolution,  we exploit HVA with shallow or moderate depths. There are two reasons: firstly, HVA may be viewed as a shortcut for QAA by trotterization; secondly, an accurate solution is not demanded since we just need to exploit the asymmetry of the algorithm.  For this, we propose a variational quantum algorithm, called Delta-VQE, which aims to reveal the asymmetry.  It compares two optimized energies for a Hamiltonian that uses two different reference states. The quantum algorithmic procedure of Delta-VQE can be described as following:
\begin{enumerate}
	\item  Use the Hamiltonian variational ansatz of the same depth $p$ to solve a given Hamiltonian $H(h)$ with two different reference states ($\ket{\psi_0}$ and $\ket{\psi_1}$) corresponding to two distinct phases of matter. The optimized energy is denoted respectively as,
	\begin{eqnarray*}
			&&E_0(h)=\bra{\psi_0}U_0^\dagger(\boldsymbol{\theta}_0^*)H(h)U_0(\boldsymbol{\theta}_0^*)\ket{\psi_0} \nonumber \\
			&&E_1(h)=\bra{\psi_1}U_1^\dagger(\boldsymbol{\theta}_1^*)H(h)U_1(\boldsymbol{\theta}_1^*)\ket{\psi_1} \nonumber, 
	\end{eqnarray*}
	where $\boldsymbol{\theta}_0^*$ and $\boldsymbol{\theta}_1^*$ are optimized parameters corresponding to the reference states $\ket{\psi_0}$ and $\ket{\psi_1}$, respectively. In constructing HVA for two references stats, the order of Hamiltonian evolutions in Eq.~\eqref{eq:HVA} may be different, and thus $U_0$ and $U_1$ may not be the same~(see Eq.~\eqref{HVA_TFIM} for a detailed example).   
	 
	\item Calculate the absolute difference of two optimized energies 
	\begin{equation}\label{eq:Delta_E}
		\Delta E(h)=|E_0(h)-E_1(h)|,
	\end{equation}
	 and show the relation between $\Delta E(h)$ and $h$. The minimum point of $\Delta E(h)$ at $h_c$ will be identified as the quantum critical point.  
\end{enumerate} 

 By construction, $\Delta E(h)$ can be small for $h\neq h_c$ if the depth $p$ is large enough that VQE can solve the ground state accurately, regardless of the choice of the initial state. Thus, a signature of quantum critical point may be not sharp if locating it as a minimum. This can suggest that we may use a small $p$ in the Hamiltonian variational ansatz. In the next section, we will numeral demonstrate the Delta-VQE algorithm using small $p$.

\section{Results}\label{sec:results}
We now test the Delta-VQE for locating quantum critical points on several typical quantum systems, the transverse field Ising model~(TFIM), the spin XZ model, and the clustering Ising model~(CIM). Those three models have distinct behaviors of quantum phase transitions. The TFIM has a $Z_2$ symmetry-breaking phase transition and only one side of the quantum critical point breaks the symmetry. The XZ model owns a $Z^x_2\times Z^z_2$ symmetry and the quantum phase transition is also symmetry breaking, but the two sides of the quantum critical point break different symmetries. On the other hand, the CIM has a novel quantum phase transition that both sides own ordered states. Our results on those representative models suggest that the Delta-VQE is useful for locating quantum critical points.  For the convenience of numeral simulations, all models use 8 qubits with periodic boundary conditions. The simulations are conducted using the opensource software ProjectQ~\cite{Steiger2018projectqopensource}.

\subsection{Transverse field Ising model}
The transverse field Ising model is a standard model for studying quantum phase transition~\cite{Sachdev}. The Hamiltonian of TFIM can be described as follows, 
\begin{equation}
	H_{\text{TFIM}}(h)=-\sum_{i=0}^{N-1} \sigma_i^z \sigma_{i+1}^z -h\sigma_i^x. 
\end{equation}
The TFIM owns a $Z^z_2$ symmetry and is invariant under $\sigma^z$ to $-\sigma^z$. The quantum critical point locates at $h=1$ and the symmetry-breaking occurs for $h<1$.

For the purpose of using the HVA ansatz, we set  $H_\text{TFIM}(h)=H_{zz}+hH_{x}$, where $H_{zz}=-\sum_{i=0}^{N-1} \sigma_i^z$ and $H_{x}=-\sum_{i=0}^{N-1}\sigma_i^x$. Two references states are chosen as ground states of $H_{zz}$ and $H_{x}$ respectively, namely $\ket{\psi_0}=\ket{0}^{\otimes N}$ and $\ket{\psi_1}=\ket{+}^{\otimes N}$
, where $\ket{+}=(\ket{0}+\ket{1})/\sqrt{2}$. It is noted that there is another choice $\ket{\psi_0}=\ket{1}^{\otimes N}$ that also breaks the $Z_2$ symmetry. 

The HVA ansatz is constructed slightly different for two reference states, 
\begin{eqnarray} \label{HVA_TFIM}
	&&\ket{\psi_0(\boldsymbol{\theta})}=\prod_{i=1}^p 
\exp(i\,\theta_{i2}H_{zz})\exp(i\,\theta_{i1}H_{x})\ket{0}^{\otimes N}\nonumber\\
	&&\ket{\psi_1(\boldsymbol{\theta})}=\prod_{i=1}^p 
	\exp(i\,\theta_{i2}H_{x})\exp(i\,\theta_{i1}H_{zz})\ket{+}^{\otimes N}.
\end{eqnarray}
The ordering of evolution of $H_{zz}$ and $H_{x}$ is different for two reference states, since evolution of $H_{zz}$~($H_{x}$) will not rotate $\ket{0}^{\otimes N}$~($\ket{+}^{\otimes N}$). This can make sure that two HVA ansatzs of depth $p$ have effective $2p$ parameters and thus the absolute energy difference can be constructed fairly. The scheme of constructing HVA ansatz for different reference states with different ordering of Hamiltonian evolution will be implemented for all models considered in this work. 

It is important to note that the parameterized wavefunction $\ket{\psi_0(\boldsymbol{\theta})}$~($\ket{\psi_1(\boldsymbol{\theta})}$) has the same symmetry as the reference state $\ket{0}^{\otimes N}$~($\ket{+}^{\otimes N}$), since the unitary operator of HVA preserves the same symmetry of the Hamiltonian. Thus, $\ket{\psi_0(\boldsymbol{\theta})}$ breaks the $Z_2^z$ symmetry while $\ket{\psi_1(\boldsymbol{\theta})}$ preserves this symmetry. This will have a significant effect on the accuracy for preparing the ground state of a Hamiltonian $H(h)$ using two different reference states, which will be discussed later.    

The optimized parameters in parameterized wavefunction  $\ket{\psi_0(\boldsymbol{\theta})}$ and $\ket{\psi_1(\boldsymbol{\theta})}$
should be determined by minimizing their variational energies of $H_\text{TFIM}(h)$, respectively. Then, the absolute energy difference $\Delta E(h)$ is obtained according to Eq.~\eqref{eq:Delta_E}. 

\begin{figure} 
	\centering
	\includegraphics[width=1\columnwidth]{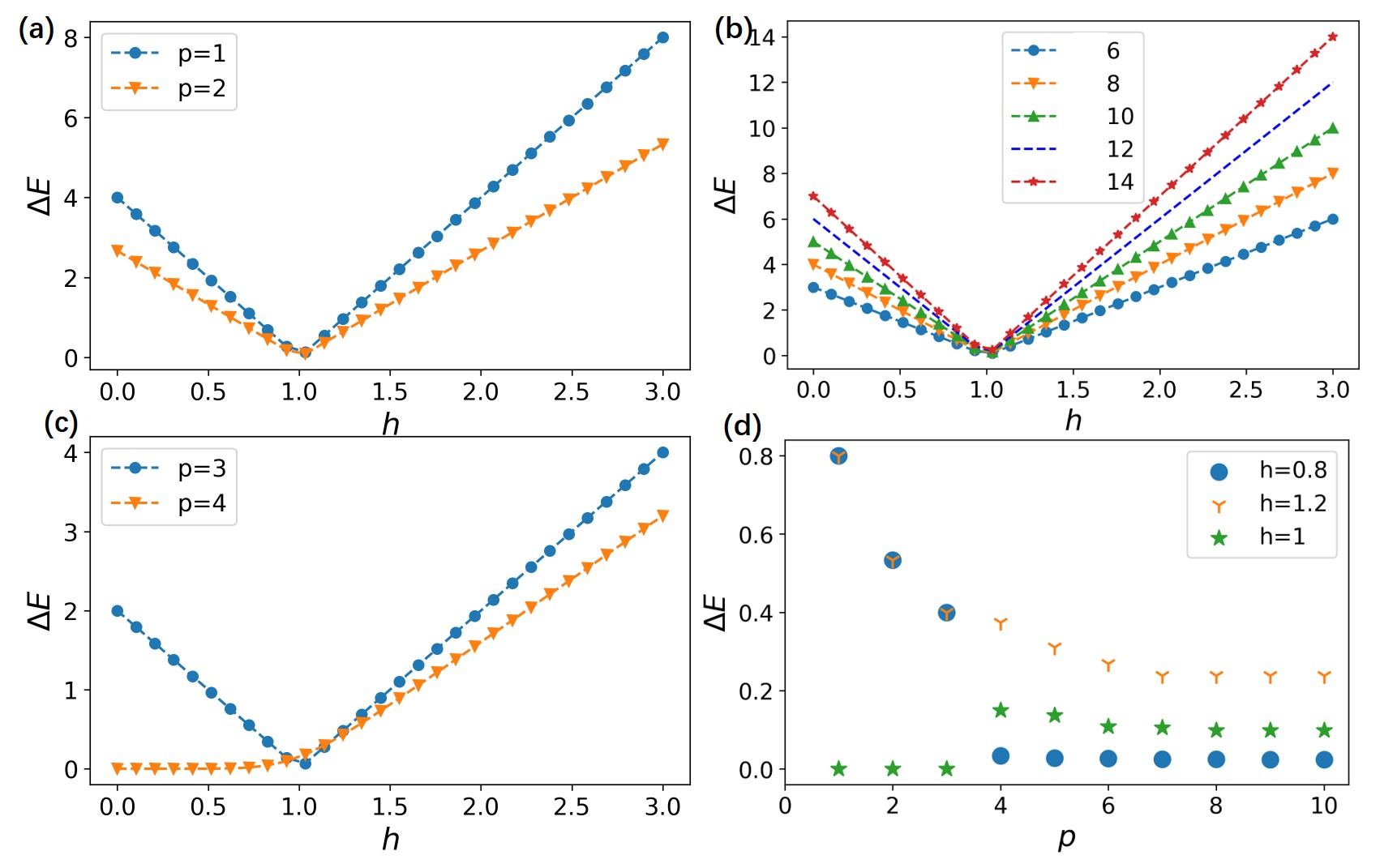}
	\caption{Delta-VQE for locating the critical point for the TFIM. (a). The absolute variational energy difference $\Delta E(h)$ shows a clear turning point at $h=1$ (shown with lattice size $N=8$); (b). $\Delta E(h)$ for lattice size from $N=6$ to $N=14$; (c). The tuning point vanishes for larger $p$, .e.g, $p=4$; (d). Behavior of $\Delta E(h)$ for different depths $p$, and $\Delta E(h=1)$ will raise from zero to nonzero when $p$ increases.}
	\label{fig:TFIM_1}
\end{figure}

In the numeral simulation, different $p$ for HVA ansatz is chosen. It can be seen in Fig.~\ref{fig:TFIM_1}a that for very small $p$, e.g., $p=1,2$, the dependence of $\Delta E(h)$ with $h$ shows a clear $V$ shape pattern, and the turning points~(the minimum point) locate at the quantum critical point $h=h_c=1$ exactly. Notably, the signature of the turning point as a quantum critical point is sharper for $p=1$. Moreover, the clear signature holds for different lattice sizes as revealed in Fig.~\ref{fig:TFIM_1}b. This suggests that the shallower circuit in the Delta-VQE can be more suitable for locating the quantum critical point. To further support the suggestion, we also evaluate $\Delta E(h)$ for $1\leq p\leq 10$ under $h=h_c$, $h<h_c$ and $h>h_c$, as seen in Fig.~\ref{fig:TFIM_1}c. At larger $p>3$, it can be seen that $\Delta E(h=0.8)$ can be smaller than $\Delta E(h=1)$, indicating $\Delta E(h)$ will not be a signature of quantum critical point. Indeed, the curve $\Delta E(h)$ will not have a sharp turning point for $p=4$(see Fig.~\ref{fig:TFIM_1}d). On the other hand, the behavior of $\Delta E(h=1)$ is distinct since it will increase for $p\geq3$, while $\Delta E(h\neq1)$ will decrease as $p$ increases. This may be considered as another signature of a critical point when there is a symmetry-breaking quantum phase transition.

The distinct behavior of $\Delta E(h=1)$ at larger depth $p$ may be understood as follows. The variational wavefunction $\ket{\psi_0(\boldsymbol{\theta})}$( $\ket{\psi_1(\boldsymbol{\theta})}$) disobeys~(obeys) the spin-flip symmetry which inheres from the reference state $\ket{\psi_0}$~($\ket{\psi_1}$). As the critical state should respect the spin-flip symmetry, it can be seen that $\ket{\psi_1(\boldsymbol{\theta})}$ can better parameterize the critical state than $\ket{\psi_0(\boldsymbol{\theta})}$. On the other hand, for very small $p$ both are very inaccurate for describing the critical state due to a limited expressive power of the circuit. This seemly shortcoming, nevertheless, can be exploited to locate the critical point by comparing the variational energies. 

\begin{figure} 
	\centering
	\includegraphics[width=0.8\columnwidth]{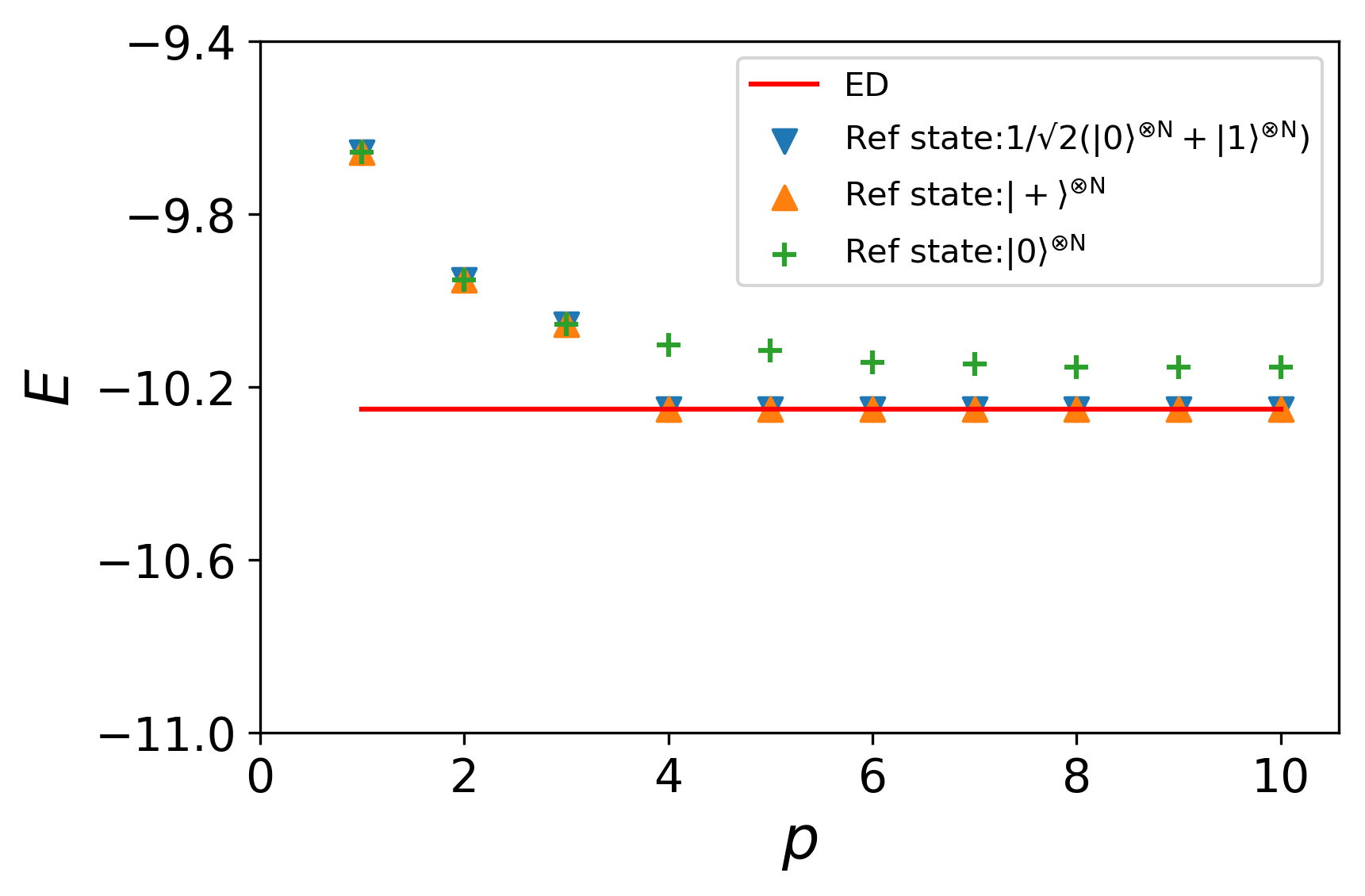}
	\caption{Optimized energies for TFIM using different references states. The state $\ket{0}^{\otimes N}$ is symmetry-breaking while the other two are symmetric with regard to $Z_2$ flip symmetry $\ket{0}\leftrightarrow\ket{1}$.}
	\label{fig:TFIM_2}
\end{figure}

To further understand the role of symmetry, we also use a symmetric reference state $\ket{\psi^s_0}=\frac{1}{\sqrt{2}}(\ket{0}^{\otimes N}+\ket{1}^{\otimes N})$ to replace $\ket{0}^{\otimes N}$. With the symmetric reference state, it can be seen that $\Delta E(h=1)$ will become zero for large $p$ again, as can be indicated in Fig.~\ref{fig:TFIM_2}
that two optimized variational energies using two symmetric states are equal.
In practice, however, preparing an entangled state $\ket{\psi^s_0}$ requires a circuit depth of the lattice size $N$ and is demanding for quantum resources. Thus, it is better to adopt two easy-to-prepare product states $\ket{\psi_0}$ and $\ket{\psi_1}$ and use a shallow circuit with small $p$. 


\subsection{Spin XZ model}
We now turn to the next typical model with a different kind of symmetry-breaking quantum phase transition than the transverse field Ising model. It is the spin XZ model~\cite{Barouch_PRA_1971,Latorre2004GroundSE}, whose Hamiltonian is, 
\begin{equation}
	H_{xz}(h)=-\sum_{i=0}^{N-1} \sigma_i^z \sigma_{i+1}^z - h\sum_{i=0}^{N-1} \sigma_i^x \sigma_{i+1}^x. 
\end{equation}
The quantum critical point locates at $h=1$. Unlike the TFIM where only one side of the quantum critical point is symmetry-breaking, the spin XZ model owns symmetry-breaking phases at both sides. However, those two symmetry-breaking phases break different symmetries. Let us denote $Z_2^x$~($Z_2^z$) as symmetry that flips $\sigma^y\rightarrow-\sigma^y,\sigma^z\rightarrow-\sigma^z$~($\sigma^y\rightarrow-\sigma^y,\sigma^x\rightarrow-\sigma^x$). Then, ground state of $H_{xz}(h<1)$ ($H_{xz}(h>1)$) breaks the $Z_2^x$ ($Z_2^z$) but preserves $Z_2^z$ ($Z_2^x$) symmetry. 

The spin XZ model can be written as $H_{xz}(h)=H_{zz}+hH_{xx}$, where $H_{aa}=-\sum_{i=0}^{N-1} \sigma_i^a \sigma_{i+1}^a$ for $a=x,z$. Two reference states are chosen as $\ket{\psi_0}=\ket{0}^{\otimes N}$ and $\ket{\psi_1}=\ket{+}^{\otimes N}$, which are symmetry-breaking states. The parameterized wavefunctions using two different reference states are constructed similar as Eq.~\eqref{HVA_TFIM},  where the ordering of Hamiltonian evolution should be different. 

\begin{figure}
	\centering
	\includegraphics[width=1\linewidth]{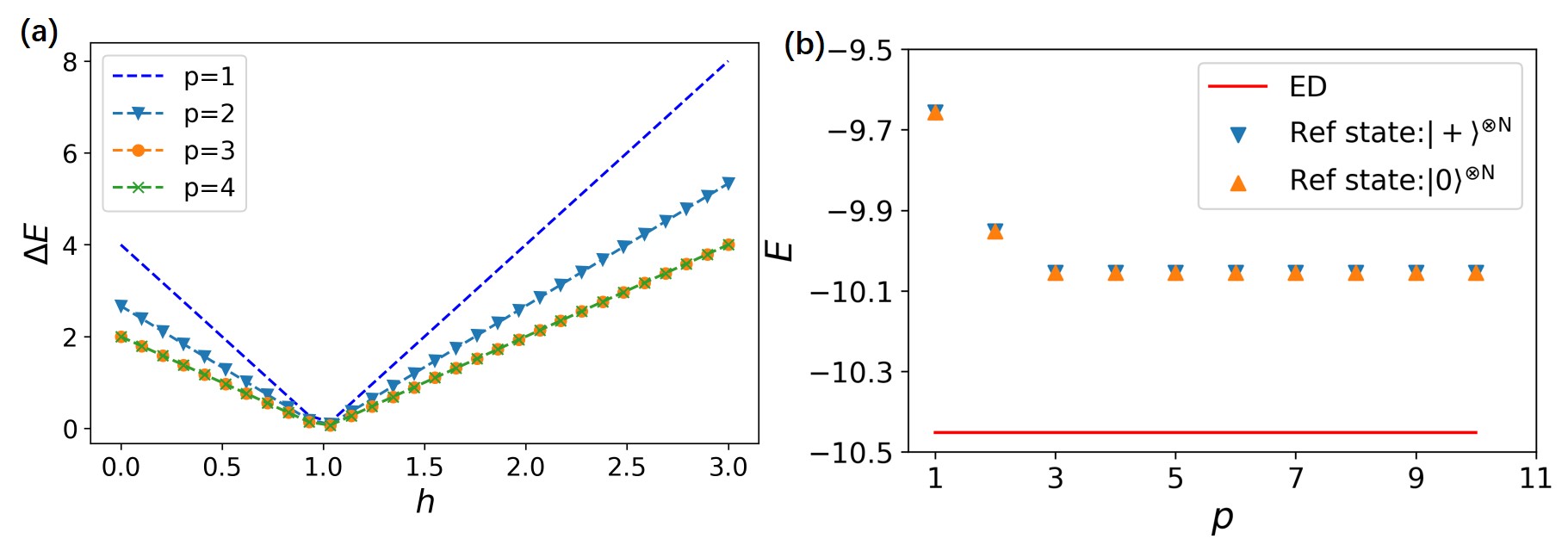}
	\caption{Delta-VQE for the spin XZ model. (a). The $\Delta E(h)$ shows a clear turning point at $h=1$ for $p=1$ and the curve converges when $p\geq3$; (b). At $h=1$, two variational energies using two different reference states converge to the same value larger that is larger than the exact ground state energy. }
	\label{fig:xzmodel}
\end{figure}

Numeral simulation of the Delta-VQE for the spin XZ model is shown in Fig.~\ref{fig:xzmodel}. As can be seen, clear shapes of $V$ with minimums locating at the critical point appear for a range of depths $p$. We further investigate optimized energies of two parameterized wavefunction for $H_{xz}(h=1)$. Both are clearly larger than the exact ground-state energy even with an increasing $p$. This can be expected since two parameterized wavefunctions will break one of two $Z_2$ symmetries and thus are limited to express the symmetric critical state faithfully. Nevertheless, the absolute energy difference is still zero, and thus can be a good signature for the critical point. 

\subsection{Cluster-Ising Model}
The third model we consider is the cluster-Ising model~\cite{Smacchia_PhysRevA_2011,Son2012TopologicalOI}, whose Hamiltonian writes as
\begin{equation}
	H_{\text{CI}}=-\sum_{i=0}^{N-1} \sigma_{i-1}^z \sigma_i^x \sigma_{i+1}^z -h\sum_{i=0}^{N-1} \sigma_i^y \sigma_{i+1}^y.
\end{equation}
The quantum critical point locates at $h=1$. 
By tuning $h$, the cluster-Ising model has a novel phase transition, from a symmetric phase with non-local string order at $h<1$ to a symmetry-breaking phase at $h>1$.

To implement the HVA ansatz, let's write $H_{\text{CI}}=H_C+hH_{y}$, where $H_C=-\sum_{i=0}^{N-1} \sigma_{i-1}^z \sigma_i^x \sigma_{i+1}^z$ and $H_y=-\sum_{i=0}^{N-1} \sigma_i^y \sigma_{i+1}^y$.
Two reference states are chosen to be ground states of $H_C$ and $H_y$, respectively. The ground state of $H_C$ is a cluster state which can be constructed as, $\ket{\psi_0}=(\prod_{i=0}^{N}CZ_{i,i+1})(H\otimes...\otimes H\ket{0}^{\otimes N}$~\cite{Smacchia_PhysRevA_2011}. Clearly, the cluster state is entangled. Moreover, it respects the $Z_2\otimes Z_2$ symmetry of the Hamiltonian. The ground state of $H_y$ can be chosen as one of the two symmetry-breaking states, e.g., $\ket{\psi_1}=\ket{+}_y^{\otimes N}$. Parameterized unitaries for two references are constructed by adjusting the ordering of $H_C$ and $H_y$, similar to the previous two models. 

\begin{figure}
	\centering
	\includegraphics[width=1\linewidth]{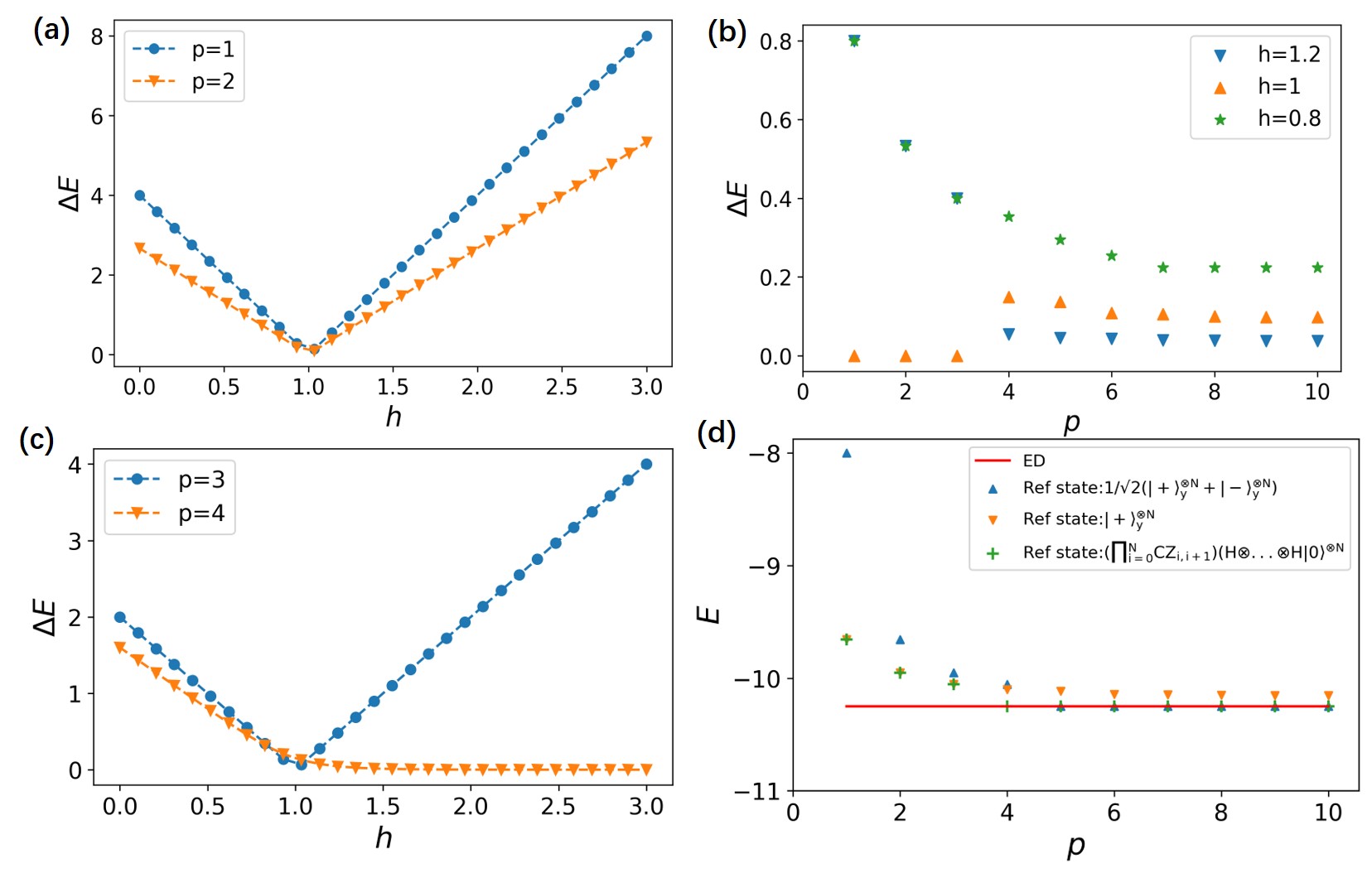}
	\caption[Cluster-Ising Model]{Delta-VQE for the Cluster-Ising model. (a). The $\Delta E(h)$ shows a clear turning point at $h=1$; (b). Distinct behavior of $\Delta E(h)$ with different depths $p$ for $h=1$;(c). The tuning point vanishes for larger $p$, .e.g, $p=4$; (d). Optimized variational energies for $h=1$ using three different reference states. }
	\label{fig:cluster-isingmodel}
\end{figure}

As seen from Fig.~\ref{fig:cluster-isingmodel}, the critical point can be revealed clearly as the turning point of the $V$-shape curve $\Delta E(h)$ for low depth $p$. However, the turning point becomes less obvious for larger $p$, e.g., $p=4$. The reason can be understood as similar to that of the TFIM system. Note that the critical state should be symmetric.  With an ansatz $\ket{\psi_0}$ using the symmetric cluster state as the initial stat, the optimized energy can converge to the exact value when $p$ increases. However, the optimized energy will not reach the exact value for the ansatz $\ket{\psi_1}$, as it is always symmetry-breaking which is deficient to express the symmetric cluster state.
On the other hand, when symmetric reference state $(\ket{+}_y^{\otimes N}+\ket{-}_y^{\otimes N})/\sqrt{2}$ is adopted, the variational energy again converges to the exact value, leading to zero $\Delta E(h=1)$. 

\section{Discussion and summary}
\label{sec:conclusion}

In summary, we have proposed a variational quantum algorithm, the Delta-VQE, to locate quantum critical points for quantum many-body systems with only shallow circuits. This can be achieved by finding the minimum of the absolute energy difference, where two optimized variational energies are obtained using two representative initial states of distinct phases of matter. We have demonstrated the usefulness of Delta-VQE for some typical quantum systems. Although each variational energy is far from accurate due to a low circuit depth, the absolute energy difference can be a good sign of the critical point. Remarkably, the signature is sharper for lower circuit depth. Our work suggests an avenue of revealing the novel properties of quantum critical point on near-term quantum computers. 

\begin{acknowledgments}
We thanks Mr.~Zhan-Hao Yuan for contributions at the early stage.
This work was supported by the National Natural Science Foundation of China (Grant No.12005065) and the Guangdong Basic and Applied Basic Research Fund (Grant No.2021A1515010317). 
\end{acknowledgments}

\bibliography{ref_delta}

\end{document}